\documentclass[pra,twocolumn,showpacs,floatfix,amsart]{revtex4-1}
\usepackage{graphicx}
\usepackage{color}
\usepackage{times,amsmath,amssymb,amstext,latexsym,float,graphicx,color}

\usepackage{ulem,float,physics}

\usepackage[hidelinks]{hyperref}

\hypersetup{colorlinks=true, citecolor=blue, urlcolor=blue, linkcolor=blue}

\usepackage{soul}

\begin{document}
 
\title{\textbf{Interplay between shell structure and trap deformation in dipolar fermi gases}}
 
\author{J. Bengtsson}
\email{jakob.bengtsson@matfys.lth.se}
\address{Mathematical Physics and NanoLund, Lund University, Box 118, 22100 Lund, Sweden}

\author{G. Eriksson}
\address{Mathematical Physics and NanoLund, Lund University, Box 118, 22100 Lund, Sweden}

\author{J.~Josefi}
\address{Mathematical Physics and NanoLund, Lund University, Box 118, 22100 Lund, Sweden}

\author{J. C. Cremon}
\address{Mathematical Physics and NanoLund, Lund University, Box 118, 22100 Lund, Sweden}

\author{S.M.~Reimann}
\address{Mathematical Physics and NanoLund, Lund University, Box 118, 22100 Lund, Sweden}
\date{\today}

\begin{abstract}
Finite fermion systems are known to exhibit shell structure in the weakly-interacting regime, as well known from atoms, nuclei, metallic clusters or even quantum dots in two dimensions. All these systems have in common that the particle interactions between electrons or nucleons are spatially isotropic.
Dipolar quantum systems as they have been realized with ultra-cold gases, 
however, are governed by an intrinsic anisotropy of the two-body interaction that depends on the orientation of the dipoles relative to each other.  
Here we investigate how this interaction anisotropy modifies the shell structure in a weakly interacting two-dimensional anisotropic harmonic trap. 
Going beyond Hartree-Fock by applying the 
so-called ``importance-truncated" configuration interaction (CI) method as well as  quadratic CI with single- and double-substitutions, 
we show how the magnetostriction in the system may be counteracted upon by a deformation of the isotropic confinement, restoring the symmetry.  
\end{abstract}

\maketitle

\section{Introduction}
Atomic alkali metal clusters are one of the first experimentally realized  man-made artificial quantum systems with mass spectra revealing  
pronounced electronic shells~\cite{deHeer1993,Brack1993} 
analogous to the closed shells in atomic noble gases, or the ``magic numbers'' of increased stability well known from nuclear structure~\cite{BohrMottelson1998}.
Another example are semiconductor quantum dots where the low-dimensional electron gas in a heterostructure can be electrostatically confined to small electron puddles exhibiting shell structure~\cite{Reimann2002}.
A third, yet very different category of finite quantum systems is offered by the advances with trapped cold atomic quantum gases
(see for example the reviews and books~\cite{Dalfovo1999,PethickSmith2002,Pitaevskii2003,Giorgini2008}). These systems offer appealing possibilities to realize and simulate a wide range of quantum many-body phenomena~\cite{Bloch2008}, for bosonic as well as for fermionic atoms. In contrast to clusters or nanostructures, there is a very high degree of tunability:  Not only can one design the shape and dimensionality of the confinement, but also modify both strength and shape of the particle interactions. It also became possible to reach the few-body limit~\cite{Sherson2010,Bakr2010,Weitenberg2011,Serwane2011},  even down to single-atom control in experiments with fermionic $^6$Li~\cite{Serwane2011}.

Initially,  studies of ultra-cold and dilute atomic quantum gases mainly considered isotropic short-range interactions~\cite{PethickSmith2002,Pitaevskii2003}. Atoms such as chromium have a large magnetic dipole moment, but usually the isotropic van der Waals interactions dominate.  However,  tuning these interactions by Feshbach resonances may enhance the relative strength of the dipole-dipole interactions~\cite{Giovanazzi2002,Werner2005},
bringing their long-range nature and spatial anisotropy into play.
After reaching the milestone of realizing a dipolar Bose-Einstein condensate~\cite{Griesmaier2005,Stuhler2005}, many 
other experiments followed~\cite{Koch2008,Lahaye2009}, also for dysprosium \cite{Lu2010,Lu2011,Kadau2016} and erbium~\cite{Aikawa2012} and polar molecules~\cite{Ni2008,Ni2010,Miranda2011,Park2015,Moses2015,Frisch2015,DeMarco2019}.  Also dipolar Fermi gases of dysprosium~\cite{Lu2012}, fermionic 
erbium~\cite{Aikawa2014}, and a chromium dipolar Fermi sea~\cite{Naylor2015} were reported. (For early reviews on dipolar gases, see, e.g., \cite{Lahaye2009,Baranov2008,Baranov2012}).  More recently, with erbium atoms it also became possible to realize two-component Fermi gases with strong and tunable interactions~\cite{Baier2018}. 

On the theory side, trapped dipolar Fermi gases have been studied extensively within variational approaches such as Hartree-Fock and beyond, see for example Refs.~\cite{Miyakawa2008,Zhang2009, Baillie2010, Zhang2010, Lima2010, Lima2010b, Ronen2010, Liu2011,Baillie2012,Baillie2012,Baillie2015,Veljic2017,Veljic2018}.
Ref.~\cite{Miyakawa2008} showed that the Fock exchange term leads to a deformation in momentum space.  The Hartree term alone resulted in a deformation in position space~\cite{Goral2001}, leading to an instability of spherically-trapped gases, but stabilizing prolate or oblate ones. The spheroidal distortion of the Fermi surface predicted in~\cite{Miyakawa2008} was  then also observed experimentally, see~\cite{Aikawa2014b}.
 
In a most simple description, shell structure occurs as a consequence of the  distribution of single-particle energies that are associated with the approximation by an effective mean-field potential. The underlying principle is similar in cases where the interactions are relatively weak; in other words, when correlation effects are not too strong. A high degree of symmetry of the system initially leads to degeneracies and a ``bunchiness" of levels in the mean-field single-particle spectrum.  When the fermion number is such that a bunch of energy levels can be fully occupied and there is a gap at the Fermi surface, {\it i.e.}, when a ``shell" is filled, the system is particularly stable. 
In an open-shell scenario, however,  where the particle number is insufficient to fill the shell,  the system will undergo a breaking of symmetry to reach stability at different level fillings by lowering the degree of degeneracy. 
In self-bound fermion systems such as nuclei or metallic clusters where interactions are mainly spatially isotropic, 
this leads to the well-known Jahn-Teller shape deformations~\cite{BohrMottelson1998,Brack1993}. In dipolar quantum gases, in contrast, the dipole-dipole two-body interaction itself is spatially anisotropic and depends on the  orientation of the dipoles relative to each other.  In this case an anisotropy of the effective mean-field  may originate rather from the intrinsic structure of the two-body force than from the trap,  confining the gas by a potential with an externally determined shape.  

We here thus pose the  question, whether shell effects may occur for fermionic dipolar gases where (electric or magnetic) dipole moments may be aligned by a corresponding  external field,  and how the shell structure modifies upon a tilt of the dipole direction. 
In three dimensions, the unavoidable head-to-tail attraction would prevent the realization of  a purely repulsive system. 
We thus here restrict our analysis to  a quasi two-dimensional harmonic trap, 
where the azimuthal symmetry yields a strong shell structure despite the reduced dimensionality~\cite{Reimann2002}, but where for moderate tilt angles the otherwise dominant head-to-tail attraction can be avoided~\cite{Ni2010,Miranda2011}.  

Early studies of both energy- and density-shell structures in trapped Fermi gases were reported in Refs.~\cite{Schneider1998,Vignolo2003,Yu2005}, followed by a time-dependent Hartree-Fock analysis of shell structure in three dimensions in Ref.~\cite{Tohyama2009}. Importantly, in the latter work it was shown that quadrupole and breathing modes are less affected in the cases of closed spherical shells. 
Ref.~\cite{Veljic2017,Veljic2018,Veljic2019} more recently investigated the effect of the dipolar anisotropy on the time-of-flight expansion and on the angular dependence of the Fermi surface deformation of the ground state in relation to the trap anisotropy,  and found that the Fermi surface deforms maximally when the dipoles are tilted along the less confined trap direction. 
Here, we analyze the interplay of this effect with the otherwise predominant shell structure in the weakly interacting regime. 

The enhancement of interaction effects in two-dimensional dipolar Fermi gases,
and the importance to go beyond Hartree-Fock was discussed in Refs.
~\cite{Parish2012, Ustunel2014,Ancilotto2015}.
We  here apply the so-called ``importance-truncated" configuration interaction (CI) method~\cite{Roth2009} as well as  quadratic CI~\cite{Pople1987} with single- and double-substitutions, allowing us to go beyond Hartree-Fock. %This was found necessary especially in the open-shell cases. 

\smallskip

The paper is organized as follows: In Sec.~\ref{secSys} we describe the setup and the effective dipole-dipole interaction in a quasi two-dimensional confinement, and describe in Sec.~\ref{secMet} the methodology. 
In Sec.~\ref{secLow} we discuss the low-lying energy eigenstates as a function of the dipolar anisotropy for a deformed confinement, mapping out the interplay between  the trap confinement and the magnetostriction associated with the dipolar two-body interaction.  We finally analyze  the second differences in the ground-state energies, which is a measure for the strength of the shell structure in the symmetric, anisotropic and symmetry-restored cases, in Sec.~\ref{secEnshell}, and conclude in Sec.\ref{secConcl}.

\section{Dipoles in a quasi two-dimensional harmonic trap}
\label{secSys}

Let us now consider $N$ spin-polarized fermions in a trapping potential 
%\begin{equation}
 $ V^{\text{trap}}(\pmb{r}) = V^{\text{trap}}_z(z) V^{\text{trap}}_{\bot}(\pmb{r}_\bot)$,
%\end{equation}
where $\pmb{r}=(x,y,z)$ and $\pmb{r}_\bot = (x,y)$. Here, $V^{\text{trap}}_z$ is a tightly confined harmonic oscillator (with angular frequency $\omega_z$), such that all particles may be described by the (single-particle) Gaussian ground state in the $z$-direction, and 
%  \end{equation*}
\begin{equation}
  V^{\text{trap}}_\bot(\pmb{r}_\bot) = \frac{1}{2}M \omega^2_\bot (\alpha x^2 + \alpha^{-1}y^2) \label{Vtrap}
\end{equation}
is a parity-conserving potential, where $M$ is the particle mass, $\omega_\bot = 0.01\omega_z$ %(i.e. $\omega_\bot \ll \omega_z$)
and $\alpha = 1.15$. The constant $\alpha$ determines the anistropy of the trap in the $xy$-plane. 
\begin{figure}[ht]
  \includegraphics[width=0.9\linewidth]{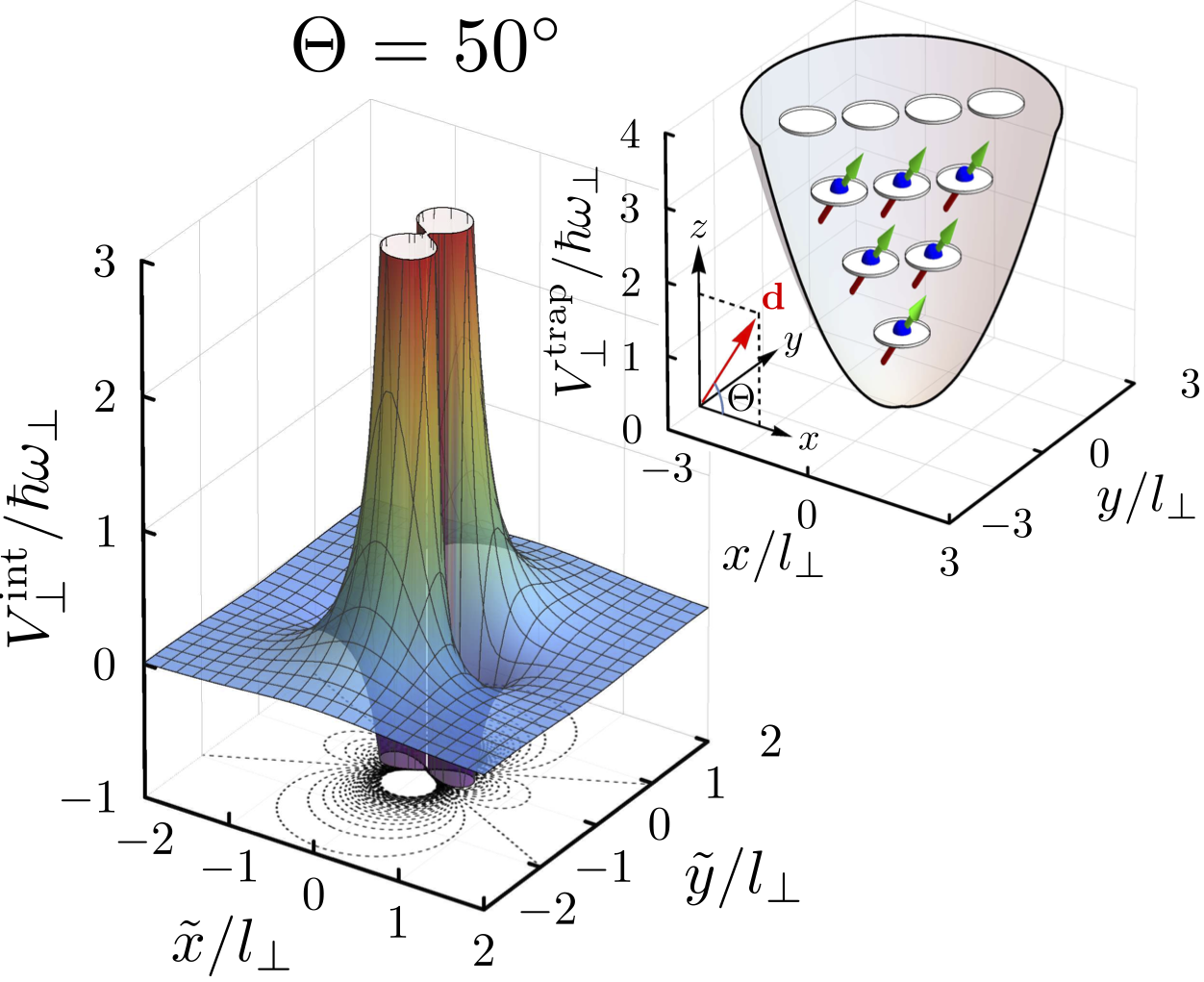}
  \caption{{\it (Color online)} Effective interaction potential $V_{\perp }^{\mathrm {int}}$, given by Eq.~\ref{Vint}, for interaction strength $D^2=0.8^2 \hbar\omega_\bot l_\bot^3$, oscillator length $l_\bot = 10l_{z}$ and a dipolar angle of $\Theta = 50^{\circ }$.
The inset  to the upper right shows a schematic picture of the quasi two-dimensional harmonic trapping potential ($\alpha=1$) and its single-particle states in the first four shells, where the white circles schematically represent the single orbitals with shell degeneracies $d_{N_0}=N_0+1$, $N_0=0,1,2\dots $ in two dimensions.}
  \label{Fig:system}
  \end{figure}
The dipole moment vectors $\pmb{d}$ associated with the fermions are aligned by an external field. The interaction potential $V^{\text{int}}$ between two particles, with identical vectors $\pmb{d}$, respectively at positions $\pmb{r}$ and  $\pmb{r}^\prime$, reads
\begin{equation}
  V^{\text{int}}(\tilde{\pmb{r}}) = D^2\frac{1 - 3 \cos^2 \theta_{\tilde{\pmb{r}}\pmb{d}}}{\lvert \tilde{\pmb{r}}\rvert^3}, \label{VintFull}
  \end{equation}
where $D^2=0.8^2 \hbar\omega_\bot l_\bot^3$ is the coupling (or interaction) strength, $l_\bot = \sqrt{\hbar/(M\omega_\bot)}$ is the oscillator length in the $xy$-plane, $\tilde{\pmb{r}} = \pmb{r}-\pmb{r}^{\prime}$ is the relative position  and $\theta_{\tilde{\pmb{r}}\pmb{d}}$ is the angle between $\tilde{\pmb{r}}$ and $\pmb{d}$. The tight confinement in the $z$-direction effectively reduces the three-dimensional potential, $V^{\text{int}}$, to a quasi two-dimensional one, $V^{\text{int}}_{\bot}$.  As in Ref.~\cite{Cremon2010} we assume that $\pmb{d}$ is oriented in the $xz$-plane by an angle $\Theta$ to the x-axis, and obtain
\begin{align}
 V^{\text{int}}_\bot(\tilde{\pmb{r}}_\bot) & = 
  \frac{D^2e^{\tilde{\xi}}}{2\sqrt{2\pi} l^3_z} \bigg \{ 
  (2+4 \tilde{\xi}) K_0({\tilde{\xi}}) -4 \tilde{\xi} K_1({\tilde{\xi}}) \nonumber \\ & + \cos^2 \Theta \left[ -(3+4\tilde{\xi}) K_0(\tilde{\xi}) + (1+4\tilde{\xi}) K_1(\tilde{\xi}) \right] \nonumber \\ & + 2\cos^2 \Theta \cos^2 \tilde{\phi} \left[ -2\tilde{\xi} K_0(\tilde{\xi}) + (2\tilde{\xi}-1) K_1(\tilde{\xi}) \right] \bigg \}  \label{Vint} \nonumber \\
\end{align}
where  $\tilde{\xi} = \tilde{r}^2_\bot / (4 l^2_z) = (\tilde{x}^2+\tilde{y}^2)/(4 l^2_z)$,  $l_z = \sqrt{\hbar/(M\omega_z)}$ is the oscillator length in the $z$-direction, $\tilde{\phi} = \arctan({\tilde{y}/\tilde{x}})$ and where $K_0$ and $K_1$ are modified Bessel functions of the second kind.  The effective dipole-dipole interaction is spatially isotropic when $\Theta = 90^\circ$, but for smaller angles $\Theta $, it has a pronounced spatial anisotropy, as sketched in~Fig.~\ref{Fig:system} for $\Theta =50^{\circ} $. (The inset shows the shell degeneracies in a two-dimensional isotropic harmonic trap).  
The effective Hamiltonian of the system here thus reads  
\begin{equation}
\hat{H} = \sum_{i=1}^N \left[ \frac{\hat{\pmb{p}}^2_{\bot,i}}{2M} + V^{\text{trap}}_\bot(\hat{\pmb{r}}_{\bot,i}) \right] + \sum_{i > j}^N V^{\text{int}}_\bot(\hat{\pmb{r}}_{\bot,i} -\hat{\pmb{r}}_{\bot,j}),  \label{H}
\end{equation}
where $V^{\text{trap}}_\bot$ was given in Eq.~(\ref{Vtrap}) above, $V^{\text{int}}_\bot$ in Eq.~(\ref{Vint}), and  $\hat{\pmb{p}}_{\bot,i}=(\hat{p}_{i,x},\hat{p}_{i,y})$ and $\hat{\pmb{r}}_{\bot,i}=(\hat{x}_i,\hat{y}_i)$ are the momentum- respectively the position operator of the $i$:th particle in the $xy$-plane. For the considered system, we observe that parity is preserved. The individual many-body eigenstate to $\hat{H}$ has thus either even or odd parity.

\section{Going beyond Hartree-Fock}
\label{secMet}
We now search for the low energy eigenstates to $\hat{H}$ in Eq.~(\ref{H}) for $N=1,2,\ldots, N_{\text{max}}$ fermions and for different dipolar orientations $\Theta$. Depending on $N_{\text{max}}$, two different numerical methods are employed for $N > 1$; so-called importance-truncated configuration interaction (ITCI)~\cite{Roth2009} and quadratic configuration interaction with single and double substitutions (QCI-SD)~\cite{Pople1987}. (For $N=1$, the solution is trivially known from  the single-particle oscillator ground state.) In all methods, the same primitive one-body basis 
\begin{equation}
\varphi_{m,j}(r_\bot, \phi) =  e^{im \phi}B^{(k)}_{j}(r_\bot)
  \end{equation}
is used at the most fundamental level, where $(r_\bot,\phi)$ are the polar coordinates,  $m$ is an integer and $B^{(k)}_{j}$ is a $k$:th order B\nobreakdash-spline. The B\nobreakdash-splines are piece-wise polynomials defined by their order $k$ and by their so-called knot-point sequence $\tau_{j} \leq \tau_{j+1}$, see e.g. Ref.~\cite{DeBoor1978}, 
\begin{align}
  B_{j}^{(1)}(r_\bot) =& \left\{ \begin{array}{ll} 1 & \text{ if } \tau_j \leq r_\bot < \tau_{j+1} \\ 0 & \text{ otherwise} \end{array} \right. \\
  B_{j}^{(k)}(r_\bot) =& \frac{r_\bot -\tau_j}{\tau_{j+k-1}-\tau_j} B^{(k-1)}_{j}(r_\bot) \nonumber \\ &+ \frac{\tau_{j+k}-r_\bot}{\tau_{j+k}-\tau_{j+1}} B^{(k-1)}_{j+1}(r_\bot) 
\end{align}
In total, we use 702 single-particle basis states. The angular part of $\varphi$ is limited to $m \in [-13,13]$ and 26 B\nobreakdash-splines are used to sample the radial part. We chose fifth order B\nobreakdash-splines, i.e. $k=5$, defined by a linear distribution of knot-points with $\Delta \tau = 0.15 l_\bot$ in an inner region, $0 \leq r_{\bot}/l_\bot \leq 3 $, and an exponentially increasing distance between the knot points in an outer region, $3 \leq r_{\bot}/l_\bot \leq 5$. %Here, $l_\bot = \sqrt{\hbar/(M\omega_\bot)}$ is the oscillator length in the $xy$-plane.
The last knot point, located at $r_\bot = 5 l_\bot$, sets the radius of our computational box. 

For a systematic comparison of systems with only a handful of fermions, the eigenergies and eigenstates of $\hat{H}$ may, in principle, be retrieved individually using ITCI. Here, we divide the full many-body Hilbert space $\mathcal{H}$ into  a model (or reference) subspace $\mathcal{H}_M$ and the orthogonal complementary $\mathcal{H}_C$, where $\mathcal{H}_M$ is tailored for a specific many-body solution at a time. To reduce the computational workload, the diagonalization of $\hat{H}$ is performed within $\mathcal{H}_M$ alone. The validity of the model space, for the targeted many-body solution, is subsequently estimated using perturbation theory. In particular, we expand the desired many-body solution (retrieved in $\mathcal{H}_M$) to $\mathcal{H}_C$ following the prescription of multiconfigurational first-order perturbation theory (here based on a Epstein-Nesbet like~\cite{Epstein1926,Nesbet1955} partitioning, as discussed in, e.g., Ref.~\cite{Roth2009}). The corresponding correction to the many-body eigenenergy (obtained with second order perturbation theory) serves as an error estimate of $\mathcal{H}_M$. Of course, the perturbative correction to the many-body solution also identifies the most important basis states belonging to $\mathcal{H}_C$. This information promotes an iterative construction of $\mathcal{H}_M$,  starting from an initial guess spanned by only a few many-body basis-states, and transferring the most important states from $\mathcal{H}_C$ to $\mathcal{H}_M$ in each iteration. The incorporation of new basis states is repeated until the model space becomes adequate, with an error estimate below some pre-defined threshold. If the final estimate of the error is small, then the computed eigenstates and eigenenergies are expected to be close to those obtained with full configuration interaction. Since the latter method is size-extensive, this permits a direct comparison between states of contrasting particle numbers. In general, however, the rapid increase in the size of the Hilbert space with $N$ effectively puts a limit to the applicability of the considered method. With 702 one-body basis states, there are for instance already $\sim 1.00\times 10^{10}$ possible many-body states to consider for $N=4$ (even though not all of them have the desired parity and are deemed important enough to belong to $\mathcal{H}_M$).

For slightly larger particle numbers, we instead make use of the method of quadratic configuration interaction with single- and double substitutions, QCI-SD. Here, the mean-field (Hartree-Fock) orbitals are first retrieved in the $\varphi$\nobreakdash-basis. Next, to go beyond the mean-field descriptions, we compute (in a self-consistent procedure) the contributions connected to the many-body states constructed from the mean-field solution by single- and double excitations, while at the same time enforcing size-extensivity. In a way, the considered method can be seen as a modification of the configuration interaction with single and double substitutions, making it size extensive. Alternatively, QCI-SD may be thought of as a simplified version of the corresponding coupled cluster method with single- and double substitutions. With QCI-SD and given the considered one-body basis size, only $\sim1.46\times 10^6$ many-body basis-states are, e.g., referenced for $N=4$ and $\sim2.82\times 10^7$ states for $N = 16$.  
To speed up (and in some cases to facilitate) the convergence of the self-consistent QCI-SD calculation, the frequently used direct inversion of the iterative subspace method, DIIS~\cite{Pulay1980}, is here implemented.
Note also that, in order to construct the many-body low energy spectrum,  in addition QCI-SD is  here applied to the excited Hartree-Fock solutions, converging to local minima in the energy surface. Even when only the many-body ground state is desired, for systems where the many-body ground- and first excited states are close to one another in energy, we stress the necessity of also considering the excited Hartree-Fock solutions. When correcting for the correlation energy, an excited mean-field state may produce the lowest energy, i.e. an excited Hartree-Fock solution could be the mean-field approximation of the actual many-body ground state. To find the excited mean-field solutions, we here start the self-consistent iteration from different excitations of the mean-field ground-state and use the method of maximum overlap~\cite{Gilbert2008} together with DIIS to support the desired convergences. Alternatively, in some cases, we instead map out the different mean-field (excited) solutions by following them individually during a numerical sweep in the dipolar orientation $\Theta$. A converged mean-field solution at one angle, $\Theta$, is thus used as the initial guess for the subsequent solution at a slightly larger, or smaller, angle.

\section{Low-energy eigenstates}
\label{secLow}
Let us now first examine the properties of the ground- and the first excited state in the case of a few dipolar fermions, with interaction strength $D^2=0.8^2 \hbar\omega_\bot l_\bot^3$, confined in a slightly anisotropic trap that is elongated in the $y$-direction ($\alpha = 1.15$). For $N=1$, the ground- and the first excited state energies are $E_0/(\hbar\omega_\bot) \approx 1.0024$ and $E_1/(\hbar \omega_\bot) \approx 1.9349$ respectively. Recalling the corresponding energies $E_0/(\hbar\omega_\bot) = 1$ and $E_1/(\hbar\omega_\bot) = 2$ for an isotropic harmonic trap, we here note that the anistropic deformation shifts the two single-particle energies in opposite directions, bringing them closer to one another. For larger particle numbers, the energy structure becomes more complex (owing to the interaction between the particles) and depends explicitly on the dipolar orientation angle $\Theta$~\cite{Cremon2010}.
\begin{figure}[ht]
  \includegraphics[width=0.9\linewidth]{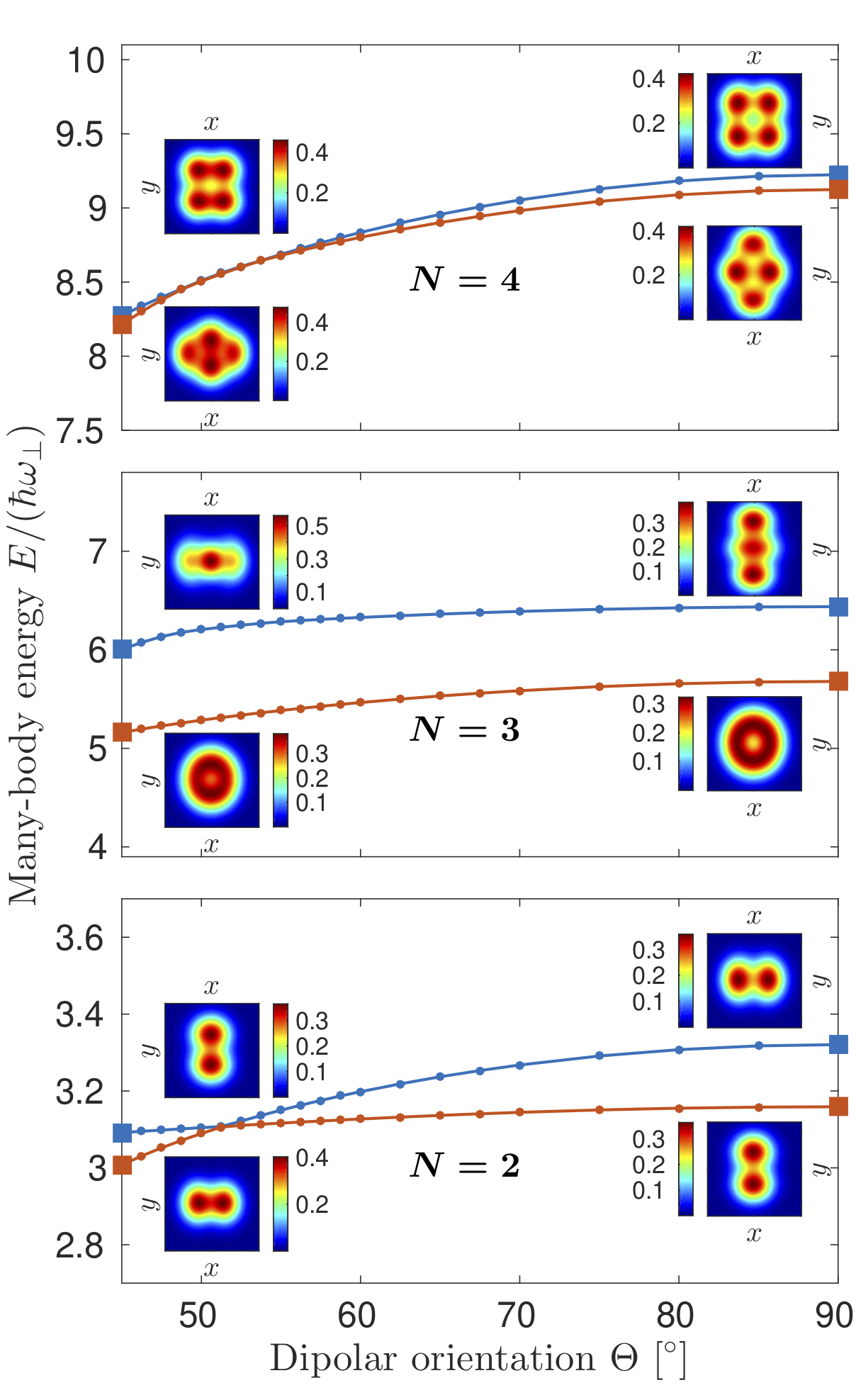}
  \caption{(Color online)  The many-body energy associated with the ground state (red line) and the first excited  state (blue line) for $2\leq N \leq 4$ and for $45^\circ \leq \Theta \leq 90^\circ$. The insets show the single-particle densities, in the interval $\lvert x/l_\bot\rvert \leq 2.5$ and $\lvert y/l_\bot\rvert \leq 2.5$, for the ground- and first-excited states at $\Theta = 45^\circ$ and $\Theta = 90^\circ$. }
  \label{Fig1}
  \end{figure}
In Fig.~\ref{Fig1}, we show the ground- and first-excited many-body energies retrieved with the ITCI method, described in Sec.~\ref{secMet} above, for $N=2,3$ and $4$, and a trap deformation of strength $\alpha =1.15$. The estimated relative error $\lvert \delta E^{(2)}_i / E_{i}\rvert < 2.5\times10^{-5}$, where $\delta E^{(2)}_i$ is the second order energy correction obtained with perturbation theory for the $i$th many-body eigenstate. (The first order correction $\delta E^{(1)}_i = 0$ by construction.) In Fig.~\ref{Fig1} we also show the equidensity distributions of the two-dimensional single-particle densities $\rho (x,y) $ for the many-body ground state at $\Theta = 45^\circ$ and $\Theta = 90^\circ$ as insets. For $\Theta = 90^\circ$, the isotropic repulsive dipole-dipole interaction results in ground-state densities that are shaped by the trap geometry. The larger extension of $\rho$ in the $y$-direction (when $\Theta=90^\circ$) is clearly seen for $N=2$ and $N=4$, but also holds for $N=3$. Upon reduction of $\Theta $, the anisotropy of the interaction potential increases. In particular, the effective dipole-dipole interaction becomes attractive for configurations where the particles are oriented ``head-to-tail'', i.e. for particles lined-up in the $x$-direction. With the decrease in $\Theta$ follows thus a possible decrease of the effective interaction energy, and consequently of the energy of the many-body ground- and low-lying states. Indeed, for $\Theta=45^\circ$ and for $N=2$ and $N=4$, we see that the ground-state densities are elongated in the $x$-direction instead, favoring ``head-to-tail'' configurations. The drastic change in ground-state density profiles are attributed to the (avoided) crossings seen around $\Theta \approx 50^\circ$ for $N=2$ and $N=4$. Actually, for $N=4$, there are two such crossings in the vicinity of $\Theta \approx 50^\circ$ (although not visible in Fig.~\ref{Fig1}), which explains the different structure in $\rho$ between the first excited state at $\Theta=90^\circ$ and the ground state at $\Theta=45^\circ$. For $N=3$, the corresponding excitation energy is larger and the change in the ground-state density distribution with $\Theta$ is less pronounced. However, the density profile of the first excited state changes greatly. Once again, (avoided) crossings are (although not shown in Fig.~\ref{Fig1}) responsible for this particular change in $\rho$, making the density elongated in the $x$-direction at lower $\Theta$. Intriguingly, for the ground state, a similar decrease in $\Theta$ triggers a fundamentally different response in $\rho$. First of all, a closer look at the ground state density reveals a minor increase in $\rho$ at the center of the trap for $\Theta=45^\circ$ compared to that for $\Theta=90^\circ$. In addition, a slightly lower value of $\langle x^2-y^2\rangle/\langle x^2+y^2 \rangle$  is found numerically for $\Theta=45^\circ$ compared to for $\Theta=90^\circ$, where $\langle x^2 \pm y^2 \rangle = \iint (x^2\pm y^2)\rho(x,y)dxdy$. Even though hardly visible in Fig.~\ref{Fig1}, a decrease from $\Theta=90^\circ$ to $\Theta = 45^\circ$ thus effectively increases the relative elongation of the ground state density in the $y$-direction, i.e. in the orthogonal direction to what is favored by the anisotropic dipole-dipole interaction.

\begin{figure}[ht]
  \includegraphics[width=0.9\linewidth]{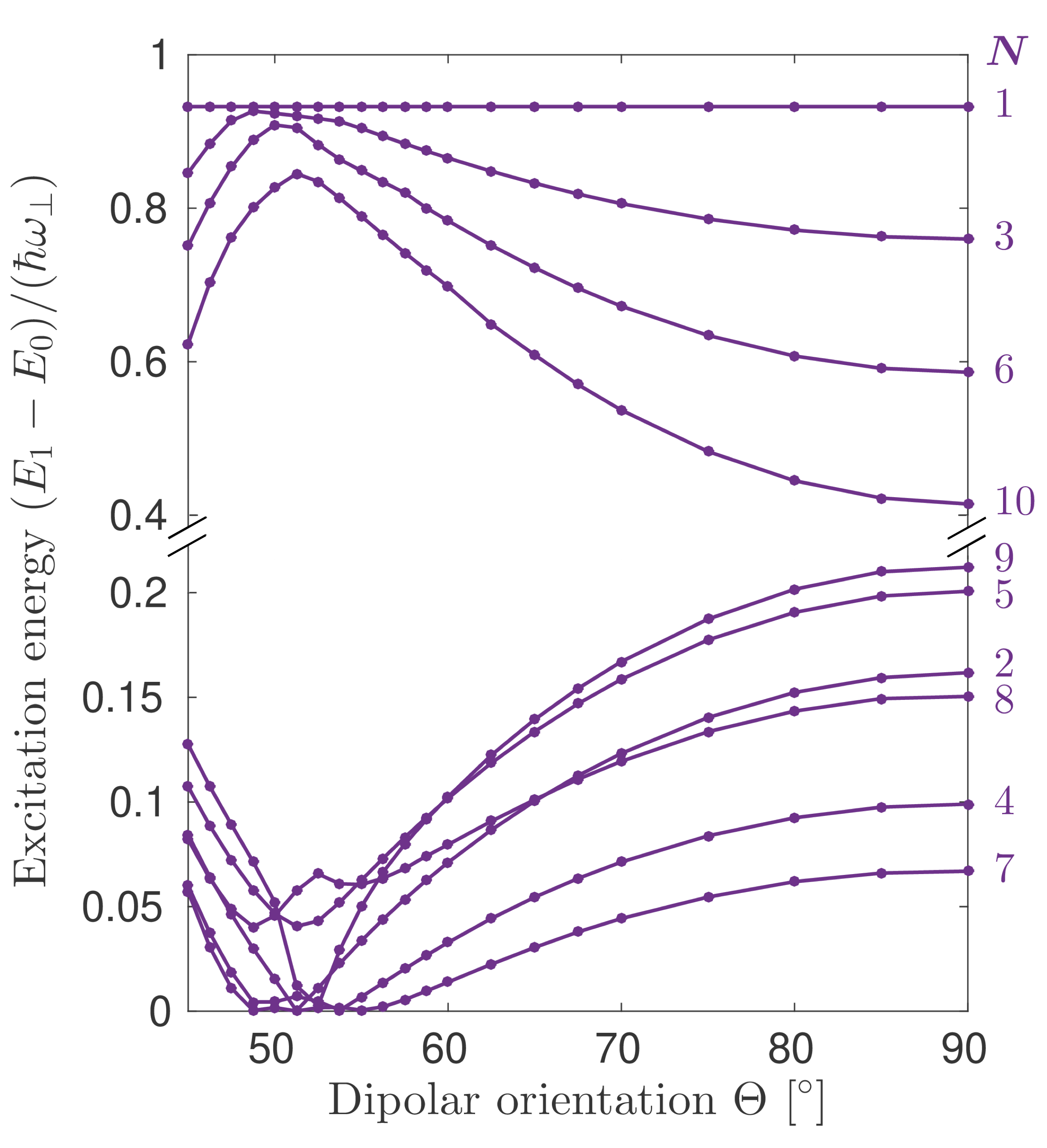}
  \caption{The energy difference $E_1-E_0$, between the first excited- and the ground state of the system, for $N \leq 10$. (Note the change of scale for the energies in the upper and lower part of the Figure). For the given values of the dipolar orientation $\Theta $, the gaps are largest in the closed-shell configurations with $N=1,3,6$ and $15$.}
  \label{Fig2}
  \end{figure}
Let us now increase the number of fermions $N$ to investigate if the system responds in a similar manner to a change in $\Theta$. In Fig.~\ref{Fig2}, the lowest excitation energies, $E_1-E_0$, retrieved with the method of QCI-SD (as discussed in Sec.~\ref{secMet}), are shown for $N\leq 10$. In general, we observe two different categories of many-body systems; those  with large excitation energies (similar for $N=3, 6$ and $10$), and those with lower energy differences (similar for $N=2, 4, 5, 7, 8$ and $9$). We also note  that the latter systems all have a minima in the excitation energy $E_1-E_0$ around $\Theta \approx 50^\circ$.
 
We recall that the many-body ground-state is  characterized by the intricate interplay between the minimization of the (dipole-dipole) interaction energy and that of the single-particle energy dictated by the trapping potential. Furthermore, for certain $N$, an (avoided) energy crossing marks the transition from the single-particle density elongated in the $y$-direction (as favored by the trap) to one elongated in the $x$-direction (as favored by the dipole-dipole interaction when $\Theta < 90^\circ$). For a larger interaction strength $D^2$ (not shown here),  the crossing shifts closer to $\Theta=90^\circ$, {\it i.e.}, a stronger contribution from the dipole-dipole interaction causes the system to respond 
already at smaller differences of the tilt angle from the  
isotropic case $\Theta = 90^\circ$. Similarly,  a larger trap deformation ($\alpha > 1.15$) shifts the crossing further away from the value $\Theta=90^\circ$.

\section{Energy shell-structure}
\label{secEnshell}

For comparison we now first briefly discuss the trivial case of non-interacting fermions in an isotropic harmonic confinement, {\it i.e.,} a system where $D^2=0$ and $\alpha=1$. In this case, the system displays a pronounced energy shell structure, trivially reflecting the azimuthal symmetry of the one-body problem that leads to the degeneracies $d_{N_0}=N_0+1$  with shell index $N_0=0,1,2,\dots $. The $N$-body ground state adheres the aufbau principle, with fermions occupying the $N$ single-particle states of lowest energy. For spin-polarized fermions, closed energy shell systems are consequently found for $N=1,3,6,10,15,\ldots $ in agreement with the single-particle orbital degeneracy  and with the Pauli exclusion principle.  

Weakly interacting trapped fermions may exhibit energy shells reminiscent to those of the non-interacting particles if the states are not too strongly correlated, which has been observed for example also in quasi two-dimensional and azimuthally symmetric  quantum dots (see the review~\cite{Reimann2002}). 
For dipolar fermions in a isotropic harmonic trap, similar energy shells are thus to be expected in the purely repulsive case at $\Theta = 90^\circ$ where the isotropic repulsion between particles renders the many-body problem rotationally invariant about the $z$-axis. The particle interaction however makes the description of distinct energy shells an approximate one, applicable only in a mean field picture: The single-particle mean-field energies, although not necessarily being degenerate, may cluster around certain energy values, {\it i.e.} in different energy shells, with larger energy gaps in between them. For a fixed confinement, the interaction energy contribution to the many-body energy grows with $N$, typically reducing the resemblance between the interacting and the non-interacting system.
\begin{figure}[h]
  \includegraphics[width=0.9\linewidth]{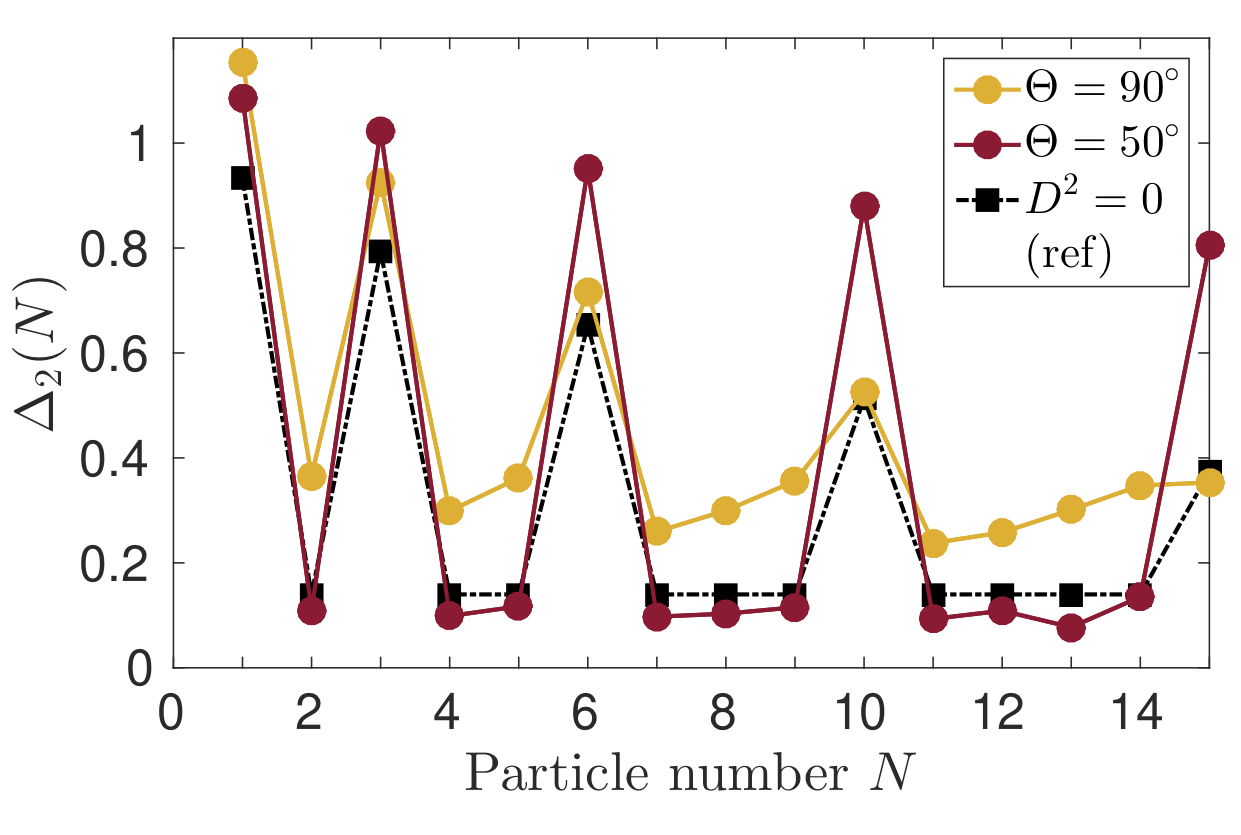}
  \caption{{\it (Color online)} The difference in addition energies, $\Delta_2(N)$, for dipolar fermions in the anisotropic confinement $\alpha = 1.15$. Two different dipolar orientations are considered; $\Theta=90^{\circ}$ (yellow line) and $\Theta = 50^\circ$ (red line) with interaction strength $D^2=0.8^2\hbar\omega_\bot l_\bot^3$. In addition, the case of non-interacting fermions ($D^2=0$, black dotted line) is shown as a reference.}
\label{Fig3}
\end{figure}

Let us now return to the case of dipolar fermions with interaction strength $D^2=0.8^2 \hbar\omega_\bot l_\bot^3$  in an anisotropic trap with $\alpha=1.15$. As a consequence of the trap deformation, the energy shell structure for $\Theta=90^\circ$ will (at least partly) be destroyed. As the single-particle orbital degeneracies associated with the isotropic harmonic oscillator are lifted, the energy shell structure is distorted already for non-interacting fermions. A reduction in $\Theta$  renders the effective dipole-dipole interaction anisotropic, i.e. also the rotational symmetry of the {\it two-body} part of the Hamiltonian gets broken. In the special case of a isotropic harmonic trap ($\alpha=1$), such  symmetry breaking necessarily weakens the energy shell resemblance for the system. For the deformed trap, however, the effect of changing $\Theta$ is more complex. The anisotropic $V_\bot$ can here, in principle, counteract the effects of the trap deformation and largely restore the energy shell structure. %Indeed, as discussed in Sec.~\ref{secLow}, a trap where $\alpha >1$ favors a single particle density $\rho$ elongated in the $y$-direction. In contrast, the dipole-dipole interaction favors a density extended in the $x$-direction (for $0 \leq \Theta < 90^\circ$). Hence, for a particular angle $\Theta$ a nearly symmetric density distribution may be found despite the otherwise dominating magnetostriction.
Intriguingly, this revival of shell structure originates from the elaborate interplay between one-body physics (governed by the anistropic trap) and two-body physics (where the anisotropy is an intrinsic property of the dipolar interaction).

For systems exhibiting a clear energy shell structure, the many-body ground states are (nearly) degenerate for the open shell systems. In other words, the lowest excitation energies observed for the open-shell systems are significantly lower than the corresponding ones for the closed-shell systems. In Fig.~\ref{Fig2}, such a feature in $E_1-E_0$ is observed for $\Theta \approx 50^\circ$ which hints  at closed energy shells for $N=1,3,6,10,\ldots$, similar to the case of non-interacting spin-polarized fermions in a two-dimensional isotropic harmonic confinement.
\begin{figure}[h]
\includegraphics[width=0.9\linewidth]{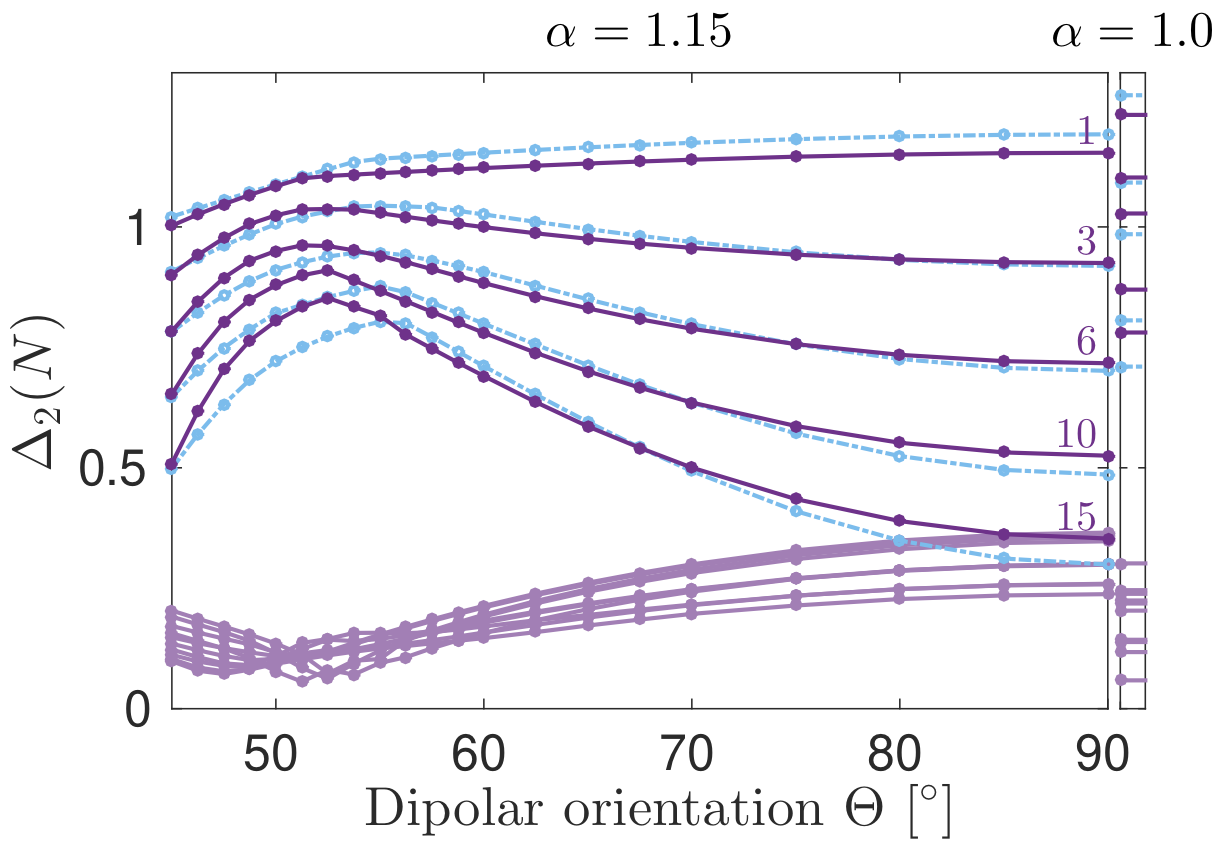}
\caption{{\it (Color online)} Addition energy differences as a function of $\Theta$ for fermions in a deformed trap with $\alpha =1.15$ (left panel) and, as a reference, for fermions with fixed dipolar orientaion, $\Theta = 90^\circ$, in a isotropic harmonic confinement, $\alpha = 1$ (right panel). In both cases, the interaction strength is $D^2=0.8^2\hbar\omega_\bot l_\bot^3$. The purple solid lines are obtained using QCI-SD, whereas the blue dashed lines  corresponds to $\Delta_2$ computed directly from the Hartree-Fock energies for $N=1,3,6,10, 15$. The purple lines associated with closed-shell systems are labeled by their particle numbers $N$ for $\alpha = 1.15$. Although not shown, for $\alpha = 1.0$, the descending order of $\Delta_2$ is also given by $N=1,3,6,10,15$. }
\label{Fig4}
\end{figure}
To characterize the shell structure in the different many-body systems characterized by the particle number $N$, we compute the addition energy difference~\cite{Capelle2007,Rontani2009}
\begin{equation}
  \Delta_2(N) = [E_0{(N+1)} - E_0{(N)}] - [E_0{(N)} - E_0{(N-1)}],
\end{equation}
where $E_0{(N)}$ is the energy of the $N$-fermion ground-state. 
When the shell structure is strong, one expects 
a series of pronounced peaks in $\Delta_2$ at the shell fillings 
which for a system close to a two-dimensional isotropic harmonic oscillator should occur at $N=1, 3, 6, 10, 15,\dots $, as we argued for above. 
Non-interacting fermions in a isotropic harmonic trap have $\Delta_2(N)/(\hbar \omega_\bot)=1$ for the closed shell $N$-particle systems and $\Delta_2(N)/(\hbar \omega_\bot)= 0$ for the open ones. In Fig.~\ref{Fig3}, we show the $\Delta_2$ obtained with the method of QCI-SD for $D^2=0.8^2 \hbar\omega_\bot l_\bot^3$, $N=1,2,\ldots,15$ and for $\Theta = 50^\circ$ (red solid line), i.e. for the dipolar orientation where we expect energy shells to manifest. For comparison, we also include $\Delta_2$ for the case where $\Theta = 90^\circ$ (yellow solid line) as well as for non-interacting fermions (black dotted line). Note that the latter two cases both show a rapid decline in the peak heights of $\Delta_2$ with $N$, indicating a loss of energy shell structure due to the deformation of the trap. For $\Theta = 50^\circ$, on the other hand, a remarkable revival of the energy shell structure can be seen, also showing a smaller  decrease in peak height in $\Delta_2$ with $N$. (We here recall that, for a fixed confinement, a zero decay rate is, in general, not possible due to the increased contribution from the interaction energy with $N$).

Finally, in Fig.~\ref{Fig4}, we examine the properties of $\Delta_2$ for different dipolar orientations in the moderately deformed trap with $\alpha =1.15$ 
in more detail. 
We find that the largest peak heights of $\Delta_2$ associated with the closed shell systems are indeed obtained for $\Theta \approx 50^\circ$, i.e. the dipolar orientation giving rise to the lowest excitation energies in Fig.~\ref{Fig2}.  We also note the slight shift in the peak heights in $\Delta_2$ (for $N=3,6,10,15$) towards higher $\Theta$ for larger $N$. A similar tendency can be seen in the corresponding maxima in the excitation energies in Fig.~\ref{Fig2}. We repeat, the larger number of particles increases the influence from the dipole-dipole interaction, which makes an angle closer to the isotropic $\Theta = 90^\circ$ favorable for shell-structures. In Fig.~\ref{Fig4} we also observe that correlations enhance the peaks in $\Delta_2$ making the shell-structure more pronounced. Next, we compare the shell structure retrieved for tilted dipoles ($\Theta \approx 50^\circ$) in a deformed trap ($\alpha = 1.15$) to the corresponding one for dipoles with $\Theta = 90^\circ$ in a isotropic harmonic confinement where $\alpha = 1.0$ (right panel). Clearly, for the isotropic trap, the peaks in $\Delta_2$ for closed shells are, at least for the first shells, higher than the corresponding ones for the deformed trap. On the other hand, the peak height in $\Delta_2$ decreases more rapidly with $N$ for the isotropic harmonic trap and drops already at $N=10$ below the corresponding one for $\alpha=1.15$ and $\Theta \approx 50^\circ$. In other words, tilted dipoles in a deformed trap seem to extend the shell-structure to slightly higher particle numbers. Here, the anistropy of the tilted dipole-dipole interaction effectively reduces the interaction energy (by introducing attractive regions) and thus prolongs the energy shell-structure of the system.        

\section{Conclusions}
\label{secConcl}

To summarize, we here investigated the effect of an anisotropic dipole-dipole interaction between (polarized) fermions on the shell structure in a quasi two-dimensional harmonic oscillator confinement. 
The interplay between magnetostriction caused by a tilt in the 
dipolar angle~\cite{Miyakawa2008}, and the stability of shells following a break in the rotational symmetry triggered by a quadrupole deformation of the isotropic trap, was examined. Even though both considered processes lower the symmetry of the Hamiltonian in effectively similar ways, their nature is inherently different: A deformation of the trap alters directly the one-body part of the problem, whereas a change in the interaction instead modifies the two-body physics.

We found that the underlying shell structure of the isotropic harmonic oscillator persists for modest interaction strengths and tilt angles. At open shells, however, there is a notable   deformation of the density distribution, caused by the dipolar magnetostriction~\cite{Veljic2017,Veljic2019}. However, this deformation can be counteracted upon by a corresponding quadrupolar deformation of the isotropic trap, restoring the overall symmetry of the problem, and thus leading to a revival of energy shell structure resembling the azimuthally symmetric case of aligned dipoles perpendicular to the trap plane. 

We finally note that the system considered here should be reachable with current experimental techniques. Giving an outlook, it would be interesting to investigate the modifications and break-down of shell structure for stronger interactions, and especially to study the interplay with the two-component properties of isospin in mixtures, such as recently experimentally realized with erbium~\cite{Baier2018}.  Especially in the mid-shell regime, where for weakly interacting gases Hund's rules prevail, tunable inter-spin interactions in combination with the dipolar anisotropy here offer new perspectives. Likewise, the extension to spin-orbit coupled dipolar Fermi gases~(see~\cite{Burdick2016}
and references therein) appears highly relevant.

\acknowledgements
We thank  E. \"O. Karabulut, L. H. Kristinsdottir, F. Malet Giralt, and M. Rontani for discussions at an early stage of this project. Our work was financially supported by the Swedish Research Council, the Knut and Alice Wallenberg foundation, and NanoLund. 

%\bibliography{dipshell.bib}

\end{document}